\documentclass{article}
\usepackage{amssymb}
\usepackage{amsmath}
\usepackage{graphicx}
\usepackage[T1]{fontenc}

\pdfoutput=1

\newcommand{\bege}{\begin{equation}}
\newcommand{\enge}{\end{equation}}
\newcommand{\beq}{\begin{eqnarray}}\newcommand{\benu}{\begin{enumerate}}\newcommand{\enu}{\end{enumerate}}
\newcommand{\eeq}{\end{eqnarray}}

\newcommand{\noi}{\noindent}

\begin{document}

\title{Stability of perturbed geodesics in $nD$ axisymmetric spacetimes}
\maketitle


\begin{center}
{\small C H Coimbra-Ara\'ujo and R C
Anjos\\Departamento de Engenharias e Exatas, Universidade Federal do Paran\'a (UFPR), Pioneiro, 2153, 85950-000 Palotina, PR, Brazil.}
\end{center}



\vspace{10pt}

\begin{abstract}

\noi The effect of self-gravity of a disk matter is evaluated by the simplest modes of oscillation frequencies for perturbed circular geodesics. It is plotted the radial profiles of free oscillations of an equatorial circular geodesic perturbed within the orbital plane or in the vertical direction. The calculation is carried out to geodesics of an axisymmetric $n$-dimensional spacetime. The profiles are computed by examples of disks embeded in five-dimensional or six-dimensional spacetime, where it is studied the motion of free test particles for three axisymmetric cases: (i) the Newtonian limit of a general proposed $5D$ and $6D$ axisymmetric spacetime; (ii) a simple Randall-Sundrum $5D$ spacetime; (iii) general $5D$ and $6D$ Randall-Sundrum spacetime. The equation of motion of such particles is derived and the stability study is computed for both horizontal and vertical directions, to see how extra dimensions could affect the system. In particular, we investigate a disk constructed from Miyamoto-Nagai and Chazy-Curzon with a cut parameter to generate a disk potential. Those solutions have a simple extension for extra dimensions in the case (i), and by solving vacuum Einstein field equations for a kind of Randall-Sundrum-Weyl metric in cases (ii) and (iii). We find that it is possible to compute a range of possible solutions where such perturbed geodesics are stable. Basicaly, the stable solutions appear, for the radial direction, in special cases when the system has $5D$ and in all cases when the system has $6D$; and, for the axial direction, in all cases when the system has both $5D$ or $6D$.
\end{abstract}




\section{Introduction}

A complete scheme to comprehend the dynamics of galaxies includes the study of many variables: the galactic shapes, the associated gravitational potentials and, as well, the galaxy main components such as gas, stars, dust, dark matter and maybe the central
supermassive black hole. Particularly, the orbit behavior of the galaxy stars represents an important element to derive galactic gravitational potentials. In this aspect, the simplest scenario is based on a fundamental approximation: although galaxies are composed of stars, we shall neglect the forces from individual stars and consider only the large-scale forces from the overall mass
distribution, neglecting small-scale irregularities due to individual stars or larger objects \cite{lin,bell}. 

The usual and most practical potentials to describe stellar circular orbits are the spherical, the axially symmetric and the barlike form potentials. In the simplest case, the stars are moving in a static, spherically symmectric potential. This potential is the appropriate one for globular clusters, which are usually nearly spherical. However, few galaxies are even approximately spherical. Many real galaxies actually approximate figures of revolution and many of them have their stars confined to the equatorial plane of an axisymmetric configuration \cite{djonghe,tremaine,polya}.       

In the present contribution, we investigate configurations associated with axisymmetric potentials for general $nD$ spacetimes with star orbits present only in the visible $3D$ space, focusing our calculations in the stability of perturbed stellar orbits. The motivation behind the $nD$ consideration resides in the general introduction of spacetime extradimensions in theories like superstrings \cite{strings}, braneworld gravity \cite{braneworld,braneworld0,braneworld1,braneworld2,braneworld3} and models of galaxies within a multidimensional universe \cite{coimbrasss,coimbrasss1,salucci}.  In other words, we want to answer the question ``could extradimensions affect the stability of the $3D$ orbits in the equatorial plane of a axisymmetric configuration?''. In this aspect, compactified or warped extradimensions should represent perturbations that possibly could break the stability of the system. The possible presence of extra dimensions in the universe is one of the most astounding features of string theory. Despite the strong theory formalism, extra dimensions still remain unaccessible and obliterated to experiments. Since the presence of ten or more spacetime dimensions is one of the central conditions of string theory and M theory, it is not unrealistic to say that experimental observation or constraints on the extra dimensions properties would be a major advance in science. In other hand, the lack of experimental evidences is usually explained by compactification which is the main geometric feature to explain why photons do not escape to the extra dimensions. Nevertheless, an alternative approach involves an extra dimension which is not compactified, as pointed by Randall-Sundrum (RS)~\cite{braneworld,braneworld0,braneworld1,braneworld2,braneworld3}. This extra dimension implies deviations on Newton's law of gravity at submillimetric scales, where objects may be indeed gravitating in more dimensions. The electromagnetic, weak and strong forces, as well as all the matter in the universe, would be trapped on a brane with three spatial dimensions. Only gravitons would be allowed to leave the surface and move into the full bulk, constituted by an anti-de Sitter - AdS$_5$ spacetime, as prescribed by RS models~\cite{braneworld,braneworld0}. Here are the main motivations concerning the choice of RS as the metric to be tested in the present paper.

The  stability of circular orbits in the equatorial plane can be studied both using an extension of the Rayleigh stability criterion \cite{vogt20} or a perturbative method based on inertial oscillations. In the Rayleigh criterion an orbit is perturbed by an infinitesimal change in the momentum of the test particle. Usually, the Rayleigh criterion is studied for a pure Newtonian circular orbit. On the other hand, oscillatory perturbative methods are based on the oscillations governed by the rotational restoring force and their frequencies. Such frequencies are characterized by the epicyclic frequency $\kappa$, defined by $\kappa^2=2\Omega(2\Omega + rd\Omega/dr)$, where $\Omega$ is the angular velocity of the disk rotation. In this sense, the radial distribution of $\kappa$ is of importance in determining the behavior of oscillations. Concerning this radial distribution, general relativity has important roles. Namely, in general relativity the epicyclic frequency does not increase monotonically inward in the radial direction, but rather reaches a maximum at certain radius and then falls to zero at the radius of marginally stable circular geodesic \cite{kato,semerak}. Oscillations outside the equatorial plane (vertical direction $z$) are also important and are decoupled from the radial oscillations.

Here, the stability will be investigated using the second method described above, namely the oscillatory perturbative method, for general axial configurations with $n$ dimensions. In this sense, the present work is divided as follows. In Section \ref{geo4} it is presented the equatorial circular geodesics in $4D$ axisymmetric fields, followed by the calculation of perturbations of such orbits, and the consequent investigation of their stability (Section \ref{stability4}). Section \ref{stabilitynd} is devoted to the derivation of general equations that describe perturbed orbits of a $nD$ axisymmetric configuration. Sections \ref{ex1}, \ref{ex2}, \ref{ex3} and \ref{ex4} present two examples to test the derived equation for the simplest case of a $5D$ configuration and two other more examples for $6D$ configurations. Explicitly, the following examples will be treated here: i) $3D$ orbits in the Newtonian limit case for a $nD$ axisymmetric configuration with no compactification (Section \ref{ex1} for $5D$ and Section \ref{ex3} for $6D$); (ii) $3D$ orbits for a pure Randall-Sundrum metric, with Weyl axisymmetric terms, that has a volcano barrier potential to prevent matter to get out from the visible $3D$ space (that works as a compactification) (Section \ref{ex21}); (iii) $3D$ orbits for a general Randall-Sundrum-Weyl metric with the same assumption about compactification (Section \ref{ex22} for $5D$ and Section \ref{ex4} for $6D$). In all cases, we introduce a cut method to generate a disk solution, such that the axial coordinate is transformed as $\bar{z} = |z|+c$, where $c>0$ is the cut parameter.

\section{Equatorial circular geodesics in (1+3) axisymmetric fields}\label{geo4}

As can be seen, e.g., in \cite{semerak}, rotating axisymmetric objects, in Weyl-Lewis-Papapetrou 1+3 cylindrical
coordinates $(t,\rho,\varphi,z)$, generate spacetimes described by the following
metric \cite{butter}

\begin{equation}
ds^2 = -e^{2\xi}dt^2 + \rho^2 B^2 e^{-2\xi} (d\varphi - \omega dt)^2 + e^{2\lambda - 2\xi} (d\rho ^2 + dz^2),
\end{equation}

\noindent
where $\xi, B, \omega$ and $\lambda$ are dependent functions of $\rho$ and $z$
only. The case where the source is an ideal fluid with zero
pressure (dust) is the same as to fix $B = 1$, reducing the number of
functions to three \cite{semerak98}. For a timelike worldline $x^\alpha (s)$ with four-velocity $u^\alpha=dx^\alpha/ds$ and angular velocity 
$\Omega = d\varphi/dt$, the specific angular momentum and the the specific energy with respect to the rest frame (at spatial infinity) are given by

\begin{equation}
\ell = u_\varphi = u^t(g_{t \varphi} + g_{\varphi \varphi} \Omega) = u^t \rho^2 B^2 e^{-2\xi}(\Omega - \omega), 
\end{equation}

\begin{equation}
E = -u_t = -u^t(g_{tt} + g_{t\varphi}\Omega) = u^t e^{2\xi} + \omega \ell.
\end{equation}

\noindent
Spatially circular orbits are the simplest type of worldline in a stationary axisymmetric field. It is also the most important worldline for the dynamics of astrophysical bodies. 
This particular case happens when $\rho=$const, $z$=const and
$\Omega=$const. The four-velocity is written as
\begin{equation}
u^\alpha = u^t (1,0,\Omega,0),
\end{equation}
where
\begin{equation}\label{ut}
(u^t)^{-2} = -g_{tt} - 2 g_{t\varphi}\Omega - g_{\varphi \varphi} \Omega^2 
= e^{2\xi} - \rho^2 B^2 e^{-2\xi} (\Omega - \omega)^2 = (E -\Omega \ell)^2,
\end{equation}
and the four-acceleration can be written as
\begin{equation}
a_\alpha = -\frac{1}{2} g_{\beta\gamma,\alpha} u^\beta u^\gamma \nonumber \\
 = - \frac{u^t_{,\alpha}}{u^t} + u^t \Omega_{,\alpha}\ell = u^t (E_{,\mu} -
\Omega \ell_{,\mu}).
\end{equation}
The equatorial symmetry is here defined as the reflectional symmetry with
respect to a plane placed at $z=0$. Naturally, only radial components of
$a_\alpha$ are non-null for circular orbits in the equatorial
plane. Nevertheless, there are two particular cases of the orbital angular
velocity where radial four-acceleration even vanishes:

\begin{equation}\label{kepler}
\Omega_{\pm} = \frac{-g_{t\varphi,\rho} \pm
  \sqrt{g^2_{t\varphi,\rho}-g_{tt,\rho}g_{\varphi\varphi,\rho}}}{g_{\varphi\varphi,\rho}}.
\end{equation}

Particularly, here the interest resides in static axisymmetric spacetimes,
namely Weyl spacetimes, with no dragging ($\omega = 0$) and null pressure
($B=1$), that reduces the number of unknown functions to only two,
$\xi(\rho,z)$ and $\lambda(\rho,z)$. The metric is now 
\begin{equation}\label{weyl}
ds^2 = -e^{2\xi} dt^2 + \rho^2 e^{-2\xi} d\varphi^2 + e^{2\lambda - 2\xi}(d\rho^2
+ dz^2).
\end{equation}
In this case, the energy-momentum tensor satisfies $T^\rho_\rho + T^z_z =
0$ and the function $\xi$ satifies the Laplace equation and the Keplerian
equatorial frequencies (\ref{kepler}) read
\begin{equation}\label{omegapm}
\Omega_{\pm} = \pm \frac{e^{2\xi}}{\rho}\sqrt{\frac{\rho\xi_{,\rho}}{1-\rho\xi_{,\rho}}}
\end{equation}

\noi with corresponding specific azimuthal angular momentum and specific energy respectively

\begin{equation}
\ell = \pm
\frac{\rho}{e^\xi}\sqrt{\frac{\rho\xi_{,\rho}}{1-2\rho\xi_{,\rho}}},\;\;\; E =
e^\xi \sqrt{\frac{1-\rho\xi_{,\rho}}{1-2\rho\xi_{,\rho}}}.
\end{equation}

\section{Stability of circular orbits in axisymmetric $4D$ fields}\label{stability4}
The  stability of circular orbits in the disk plane can  be
studied using an extension of the Rayleigh stability criterion
\cite{vogt20}, or a perturbative method where we assume that the disk
particles are describing equatorial circular geodesics in stationary
axisymmetric fields. A general relativistic equivalent method  to the Rayleigh criterion comes from a
perturbative method based in radial or vertical oscillations of the test
particle. In $4D$, this method is derived in \cite{coimbrasss1} and \cite{semerak}. Here we assume that a stable system is one in which the internal and external forces are such that any small perturbation results in forces that return the system to its prior state. In such manner, we are interested in investigate the stability of perturbed geodesics for axisymmetric orbits. The geodesic equation in a $4D$ pattern is
\begin{equation}\label{geo}
\ddot{x}^\alpha + \Gamma^\alpha_{\mu\nu}\dot{x}^\mu\dot{x}^\nu = 0.
\end{equation}
\noindent Here the perturbation of the geodesic equation is done performing $x^\alpha\rightarrow x^\alpha+\Delta^\alpha$ -- where
$\Delta^\alpha=(\delta t, \delta \rho, \delta \varphi, \delta z)$. Substituting this map in Eq. (\ref{geo}), we have
\begin{equation}
\ddot{x}^\alpha + \ddot{\Delta}^\alpha + \Gamma^\alpha_{\mu\nu}(x+\Delta)[\dot{x}^\mu+\dot{\Delta}^\mu][\dot{x}^\nu+\dot{\Delta}^\nu]=0,\end{equation}
\begin{equation}
\ddot{x}^\alpha + \ddot{\Delta}^\alpha + [\Gamma^\alpha_{\mu\nu}(x)+\frac{\partial}{\partial x^\beta}\Gamma^\alpha_{\mu\nu}\Delta^\beta][\dot{x}^\mu+\dot{\Delta}^\mu][\dot{x}^\nu+\dot{\Delta}^\nu]=0,
\end{equation}
\noindent Using (\ref{geo}) we isolate only the perturbative part
\begin{equation}
\ddot{\Delta}^\alpha + \Gamma^\alpha_{\mu\nu}\dot{x}^\mu\dot{\Delta}^\nu+\Gamma^\alpha_{\nu\mu}\dot{\Delta}^\mu\dot{x}^\nu+\frac{\partial}{\partial x^\beta}\Gamma^\alpha_{\mu\nu}\Delta^\beta\dot{x}^\mu\dot{x}^\nu=0,
\end{equation}
\noindent and how $\Gamma^\alpha_{\mu\nu}= \Gamma^\alpha_{\nu\mu}$ we finally derive geodesic equations for perturbations
\begin{equation}\label{perturbative4d}
\ddot{\Delta}^\alpha + 2\Gamma^\alpha_{\mu\nu}\dot{x}^\mu\dot{\Delta}^\nu+\Gamma^\alpha_{\mu\nu,\beta}\Delta^\beta\dot{x}^\mu\dot{x}^\nu=0.
\end{equation}
\noindent Taking the general 4D case where the coordinates in the plane of the disk are given by $x^\mu=(t,\rho=\mathrm{const},\varphi=\mathrm{const}+\Omega t,z=0)$, and the axisymmetric metric is $\mathrm{d}s^2=-\mathrm{e}^{2\xi}\mathrm{d}t^2+\rho^2 B^2\mathrm{e}^{-2\xi}(\mathrm{d}\varphi-\omega \mathrm{d}t)^2 + \mathrm{e}^{2\lambda - 2\xi}(\mathrm{d}\rho^2+\mathrm{d}z^2)$. Making $B=1$ and $\omega=0$, as discussed in the previous section, we have, for this case, that the non-null Christoffel symbols are $\Gamma^t_{t\rho}$, $\Gamma^\rho_{\rho\rho}$, $\Gamma^\rho_{zz}$, $\Gamma^z_{z\rho}$, $\Gamma^\rho_{tt}$, $\Gamma^\varphi_{\varphi\rho}$, $\Gamma^t_{\varphi\rho}$, $\Gamma^\rho_{\varphi t}$, $\Gamma^\rho_{\varphi\varphi}$, $\Gamma^\varphi_{t\rho}$, and four equations are derived
\begin{equation}\label{deltat}
(\ddot{\delta t})+2(\Gamma^t_{t\rho}+\Gamma^t_{\varphi\rho}\Omega)u^t(\dot{\delta \rho})=0,
\end{equation}
\begin{equation}\label{deltarho}
(\ddot{\delta \rho})+2(\Gamma^\rho_{tt}+\Gamma^\rho_{\varphi t}\Omega)u^t(\dot{\delta t})+2(\Gamma^\rho_{t\varphi}+\Gamma^\rho_{\varphi \varphi}\Omega)u^t(\dot{\delta \varphi})+(\Gamma^\rho_{tt,\rho}+2\Gamma^\rho_{t\varphi,\rho}\Omega+\Gamma^\rho_{\varphi\varphi,\rho}\Omega^2)(u^t)^2(\delta \rho)=0,
\end{equation}
\begin{equation}\label{deltavarphi}
(\ddot{\delta \varphi})+2(\Gamma^\varphi_{t\rho}+\Gamma^\varphi_{\varphi\rho}\Omega)u^t(\dot{\delta \rho})=0,
\end{equation}
\beq\label{deltaz}
(\ddot{\delta z}) +
(\Gamma^z_{tt,z}+2\Gamma^z_{t\varphi,z}\Omega+\Gamma^z_{\varphi\varphi,z}\Omega^2)(u^t)^2(\delta
z) = 0,
\eeq
\noindent where $\dot{x}^\mu=u^\mu=u^t(1,0,\Omega,0)$, and
$u^t\Omega=V_C$. $\Omega = |\Omega_\pm|$, given by Eq. (\ref{omegapm}) in the Weyl
axisymmetric configuration of Section \ref{geo4}.
From the proper frequency of the harmonic-oscillator equation (\ref{deltaz})
it is
possible to write down the following angular frequency $\tau_\perp$ with respect to radial
infinity, provided that the harmonic-oscillator equation in the vertical ($z$)
direction is decoupled from the others [and it is, see Eq. (\ref{deltaz})]:
\begin{equation}
\tau_\perp^2 = \Gamma^z_{tt,z}+2\Gamma^z_{t\varphi,z}\Omega+\Gamma^z_{\varphi\varphi,z}\Omega^2.
\end{equation}
For the axisymmetric metric presented in Section \ref{geo4}, i.e.,
developing the Christoffel symbols from metric (\ref{weyl}) and from the
Keplerian frequency (\ref{omegapm}), oscillations in $z$ direction read
\begin{equation}
\tau_\perp^2 = \frac{e^{4\xi - 2\lambda}}{1-\rho\xi_{,\rho}}\xi_{,zz}.
\end{equation}
To evaluate oscillations at the radial directions, suppose that the solutions
for $\delta t$, $\delta \rho$ and $\delta \varphi$ also have a form of
harmonic oscillations proportional to $e^{iKs}$, where $K$ is the proper
angular frequency and $s$ is the proper time. The condition for solvability of
equations (\ref{deltat})-(\ref{deltavarphi}) is

\begin{equation}\label{determinant_4d}
\det{\begin{pmatrix}
-K^2 & 2iK\Gamma^t_{t\rho}u^t & 0\\
2iK\Gamma^\rho_{tt}u^t&-K^2+\Gamma^\rho_{\alpha\beta,\rho}u^\alpha u^\beta&2iK\Gamma^\rho_{\varphi \varphi}u^t\Omega\\
0&2iK\Gamma^\varphi_{\varphi \rho}u^t\Omega&-K^2
\end{pmatrix}}=0,
\end{equation}

\section{General perturbed  motion equations in $nD$ spacetimes}\label{stabilitynd}

Now we need to work on calculations concerning $nD$ spacetimes (with $A,B,C,D = 0,...,n$), 
where the geodesic equations for perturbations should be written as
\begin{equation}\label{perturb}
\ddot{\Delta}^A + 2\Gamma^A_{BC}\dot{x}^B\dot{\Delta}^C +
\Gamma^A_{BC,D}\Delta^D\dot{x}^B\dot{x}^C=0,
\end{equation}
\noindent where $\Gamma^A_{BC}$ are the Christoffel symbols and
$\dot{x}^A$ are proper time derivatives
$\mathrm{d}x^A/\mathrm{d}s$. To find effective equations for perturbations in
terms only of visible fields, i.e., what are the effective expressions in terms of 
visible fields, it is necessary to analyze a general $nD$ metric
to derive a Lagrangian. The detailed steps on how to develop the motion equations can be seen 
in \cite{coimbra_rocha}. Here the novel results are the perturbed equations that will appear at the middle of 
the present section, with major posterior developings in Sections 5, 6, 7, 8 and 9. Also, in the end of this section, 
after derive the perturbed motion equations, we consequently calculate the epiciclic frequency and the orthogonal frequency.

The most general metric for such universe is given by
\begin{equation}\label{metric1} g(x^\alpha)=\begin{pmatrix}
g_{\alpha\beta}&|&g_{\alpha b}\\
-~-~-&  &-~-~-\\
g_{a\beta}&|&g_{ab}
\end{pmatrix},
\end{equation} \noi where $\alpha,\beta=0,\ldots,3$ and $a,b=4,\ldots, n$.
Furthermore we consider the convention to make the metric as a
function of only $3+1$ coordinates: $g_{AB} = g_{AB}(x^\alpha)$.
This metric components $g_{AB}$ contain the $3+1$ universe metric
terms $g_{\alpha\beta}$ and the extra dimensional terms $g_{ab}$, as
well as the crossed components. Eq. (\ref{metric1}) can be rewritten
for convenience as \begin{equation} g_{AB}=
g_{\alpha\beta}\delta^\alpha_A\delta^\beta_B +
g_{ab}\delta^a_A\delta^b_B+g_{\alpha
b}\delta^\alpha_A\delta^b_B+g_{a\beta}\delta^a_{A}\delta^\beta_B,\nonumber\end{equation}
\noi where $\delta^i_j$ are the Kronecker symbols. The derivatives
for such metric components are given by \beq\label{deriv1}
g_{AB,C}=&&g_{\alpha\beta,\gamma}\delta^\alpha_A\delta^\beta_B\delta^\gamma_C
+ g_{ab,\gamma}\delta^a_A\delta^b_B\delta^\gamma_C+g_{\alpha
b,\gamma}\delta^\alpha_A\delta^b_B\delta^\gamma_C+g_{a\beta,\gamma}\delta^a_{A}\delta^\beta_B\delta^\gamma_C.\eeq
The case $g_{AB}= g_{\alpha\beta}\delta^\alpha_A\delta^\beta_B+
g_{ab}\delta^a_A\delta^b_B$ is considered here, motivated by
formalisms where $g^a_\alpha=0$. The inverse metric is written as
\begin{equation} \label{metric_inv} g^{AB}=
g^{\alpha\beta}\delta_\alpha^A\delta_\beta^B+
g^{ab}\delta_a^A\delta_b^B,\end{equation} \noi and the derivatives are
straightforwardly provided by Eq. (\ref{deriv1})
\begin{equation}  \label{deriv2}
g_{AB,C}=g_{\alpha\beta,\gamma}\delta^\alpha_A\delta^\beta_B\delta^\gamma_C
+ g_{ab,\gamma}\delta^a_A\delta^b_B\delta^\gamma_C. \end{equation}
\noi Assuming that the spacetime has a connection presenting \emph{no} torsion, one yields the
following Christoffel symbols $\Gamma^A_{BC} =
\frac{1}{2}g^{AM}(g_{BM,C}+g_{CM,B}-g_{BC,M}).$
\noi Splitting this last expression by Eqs. (\ref{metric_inv}) and
(\ref{deriv2}) it reads \begin{equation}\label{chris_split} \Gamma^A_{BC}=
\Gamma^\alpha_{\beta\gamma}\delta^A_\alpha\delta^\beta_B\delta^\gamma_C+
\frac{1}{2}\left[g^{am}(g_{bm,\gamma}\delta^A_a\delta^b_B\delta^\gamma_C+g_{cm,\beta}\delta^A_a\delta^\beta_B\delta^c_C)-g^{\alpha\mu}g_{bc,\mu}\delta^A_\alpha\delta^b_B\delta^c_C\right].
\end{equation}
\noi The Ricci tensor components are
\begin{equation} R_{AB} = \partial_M \Gamma^M_{AB}
-\partial_B \Gamma^M_{AM} +
\Gamma^N_{AB} \Gamma^M_{NM} -
\Gamma^N_{AM}\Gamma^M_{NB}.\nonumber\end{equation}
Taking into account that the terms of the metric depends solely on
$x^\alpha$, the equation above reads

\begin{equation} \label{ricci_split}
R_{AB}=R_{\alpha\beta}\delta^\alpha_A\delta^\beta_B +
R_{ab}\delta^a_A\delta^b_B.
\end{equation}
The stress tensor can be derived from the conventional definition $T_{AB} := -2\frac{\delta
\mathcal{L}_M}{\delta g^{AB}} + g_{AB}\mathcal{L}_M$. Now, the Lagrangian for the gravitating test particles
 in a spacetime with extra dimensions, can be derived as \cite{coimbra_rocha}
\begin{eqnarray}\label{lagrangian}
L &=& (g_{AB}\dot{x}^A\dot{x}^B)^{1/2}= (g_{\alpha\beta}\dot{x}^\alpha\dot{x}^\beta + g_{a
b}\dot{x}^a\dot{x}^b)^{1/2},
\end{eqnarray}
\noindent where $\dot{x}^A = \mathrm{d}x^A/\mathrm{d}s$. The motion
equations come from the Euler-Lagrange expression
$
\frac{\mathrm{d}}{\mathrm{d}s}\left(\frac{\partial L}{\partial
\dot{x}^C}\right)-\frac{\partial L}{\partial x^C} = 0.
$
 As $\partial_A = \partial_\alpha\delta^\alpha_A +
\partial_a\delta^a_A$ and $g_{ab} = g_{ab}(x^\alpha)$ it follows that
\begin{eqnarray}
\frac{\partial L}{\partial x^C}&=& \frac{\partial L}{\partial
x^\gamma}\delta_C^\gamma + \frac{\partial L}{\partial
x^c}\delta_C^c,\nonumber\\
\frac{\partial L}{\partial x^\gamma} &=&
\frac{1}{2}L^{-1}(g_{\alpha\beta,\gamma}\dot{x}^\alpha\dot{x}^\beta
+ g_{ab,\gamma}\dot{x}^a\dot{x}^b),\nonumber
\end{eqnarray}
\noi and
$\frac{\partial L}{\partial x^c} = 0.$ It immediately yields
\begin{equation}
\frac{\partial L}{\partial x^C} =
\frac{1}{2}L^{-1}(g_{\alpha\beta,\gamma}\dot{x}^\alpha\dot{x}^\beta
+ g_{ab,\gamma}\dot{x}^a\dot{x}^b).\nonumber
\end{equation}
\noindent Likewise, the term
$\frac{\mathrm{d}}{\mathrm{d}s}\left(\frac{\partial L}{\partial
\dot{x}^C}\right)$ can be developed:
\beq
\frac{\partial L}{\partial \dot{x}^\gamma} =
\frac{1}{2}L^{-1}(g_{\gamma\beta}\dot{x}^\beta +
g_{\alpha\gamma}\dot{x}^\alpha) &=& L^{-1}g_{\mu\gamma}\dot{x}^\mu,\nonumber\\
\frac{\partial L}{\partial \dot{x}^c} =
\frac{1}{2}L^{-1}(g_{cb}\dot{x}^b + g_{ac}\dot{x}^a) &=&
L^{-1}g_{mc}\dot{x}^m.\nonumber
\eeq
\noindent Now
\begin{equation}
\frac{\mathrm{d}}{\mathrm{d}s}\left(\frac{\partial L}{\partial
\dot{x}^\gamma}\right)=L^{-1}\left[\left(\frac{\partial
g_{\mu\gamma}}{\partial x^\sigma}\right)\dot{x}^\sigma\dot{x}^\mu +
g_{\mu\gamma}\ddot{x}^\mu\right],\nonumber
\end{equation}
\noindent Also, one can write the integration constants
\begin{equation}\label{constant}
g_{cm}\dot{x}^m = N_c,
\end{equation}
\noindent since  $x^a$ are cyclic variables. Hence
$
\frac{\mathrm{d}}{\mathrm{d}s}\left(\frac{\partial L}{\partial
\dot{x}^c}\right)=0.
$
Inserting  the terms together, multiplying by $L g^{\mu\gamma}$ and
using (\ref{constant}) the equations of motion are derived:
\begin{equation}\label{motion}
\ddot{x}^\alpha + \Gamma^\alpha_{\mu\nu}\dot{x}^\mu\dot{x}^\nu =
\frac{1}{2}g_{ab,\gamma}g^{\alpha\gamma}N_c g^{ac} N_d g^{bd}.
\end{equation}
\noindent 
Clearly a plausible interpretations is that the extra dimensions
induce an external `force' in the system, that depends only on $g_{ab}$ and
$N_c$. 

Up to this point we developed the major steps explained in \cite{coimbra_rocha} to obtain the equations of motion.
From this moment we will derive the perturbed form of Eq. (\ref{motion}), highlighting the importance and the novelty 
of this new equation to evaluate the stability and the behavior of classical particles moving in, for example, axisimmetric 
orbits endowed with extra imprints. So, the perturbed form for Eq. (\ref{perturb}) is properly splitted
by doing $x^\alpha\rightarrow x^\alpha+\Delta^\alpha$ in Eq. (\ref{motion}), yielding 

\beq
\ddot{x}^\alpha + \ddot{\Delta}^\alpha +
\Gamma^\alpha_{\mu\nu}(x+\Delta)[\dot{x}^\mu+\dot{\Delta}^\mu][\dot{x}^\nu+\dot{\Delta}^\nu]=
V^\alpha (x+\Delta),
\eeq

\beq
\ddot{x}^\alpha + \ddot{\Delta}^\alpha +
     [\Gamma^\alpha_{\mu\nu}(x)+\frac{\partial}{\partial
         x^\beta}\Gamma^\alpha_{\mu\nu}\Delta^\beta][\dot{x}^\mu+\dot{\Delta}^\mu][\dot{x}^\nu+\dot{\Delta}^\nu]=V^\alpha
     + \frac{\partial}{\partial x^\gamma}V^\alpha \Delta^\gamma,
\eeq
where $V^\alpha = \frac{1}{2}g_{ab,\gamma}g^{\alpha\gamma}N_c g^{ac}N_d g^{bd}$ is the extradimensional signature. Using (\ref{motion}) we isolate only the perturbative part
\begin{equation}
\ddot{\Delta}^\alpha +
\Gamma^\alpha_{\mu\nu}\dot{x}^\mu\dot{\Delta}^\nu+\Gamma^\alpha_{\nu\mu}\dot{\Delta}^\mu\dot{x}^\nu+\frac{\partial}{\partial
  x^\beta}\Gamma^\alpha_{\mu\nu}\Delta^\beta\dot{x}^\mu\dot{x}^\nu -
\frac{\partial}{\partial x^\gamma}V^\alpha \Delta^\gamma = 0,
\end{equation}
\noindent where we expanded the Christoffel symbols and the metric in terms of
perturbations $\Delta^\beta$ (desconsidering second order perturbations) and
as $\Gamma^\alpha_{\mu\nu}= \Gamma^\alpha_{\nu\mu}$ we finally derive the
following equations for motion perturbations
\begin{equation}\label{perturbativend}
\ddot{\Delta}^\alpha +
2\Gamma^\alpha_{\mu\nu}\dot{x}^\mu\dot{\Delta}^\nu+\Gamma^\alpha_{\mu\nu,\beta}\Delta^\beta\dot{x}^\mu\dot{x}^\nu
- V^\alpha_{\;\;,\gamma} \Delta^\gamma =
0.
\end{equation}

\noi Note that the term $V^\alpha_{\;\;,\gamma} \Delta^\gamma$ contains all the information about perturbations that can be carried out by extradimensions. Note also that for $V^\alpha = 0$, Eq. (\ref{perturbativend}) recovers the
original $4D$ harmonic-oscillator, i.e., Eq. (\ref{perturbative4d}). In complement, Eq. (\ref{perturbativend}) actually 
contains four equations describing the perturbations of the visible field. Considering that perturbations have the form 
$\delta x \sim e^{iKs}$, consequently the equations for $\delta x^0$, $\delta x^1$ and $\delta x^2$ give the following 
condition for the solvability of the proper angular frequency $K$:

\begin{equation}\label{determinant_nMatrix}
\det{\begin{pmatrix}
-K^2&2iK[\Gamma^0_{01}+\Gamma^0_{21}\Omega]u^0&0\\
2iK[\Gamma^1_{00}+\Gamma^1_{20}\Omega]u^0&-K^2+\Gamma^1_{\alpha\beta,1}u^\alpha u^\beta-V^1_{\;\;,1}&2iK[\Gamma^1_{02}+\Gamma^1_{2 2}\Omega]u^0\\
0&2iK[\Gamma^2_{01}+\Gamma^2_{2 1}\Omega]u^0&-K^2
\end{pmatrix}}=0,
\end{equation}

\noindent where 

\begin{equation}
V^1_{\;\;,1} = \frac{1}{2}g_{ab,11}g^{11}N_c g^{ac}N_d g^{bd} + \frac{1}{2}g_{ab,1}g^{11}_{\;\;\;,1}N_c g^{ac}N_d g^{bd} + g_{ab,1}g^{11}N_c g^{ac}_{\;\;\;,1}N_d g^{bd}.
\end{equation}

\noindent The fourth perturbation equation is the same as Eq. (\ref{deltaz}). Note that if there are no crossed terms $g_{02}$ or $g_{20}$ of the metric (resulting $\Gamma^1_{02}=\Gamma^1_{20}=\Gamma^2_{01}=\Gamma^1_{02}=\Gamma^0_{21}=0$) and if there are no extradimensions (resulting $V^1_{\;\;,1}=0$), the expression (\ref{determinant_nMatrix}) becomes the same as (\ref{determinant_4d}). To evaluate oscillations at the radial directions, we calculate the determinant (\ref{determinant_nMatrix}) supposing that the solutions for $\delta t$, $\delta \rho$ and $\delta \varphi$ also have a form of harmonic oscillations proportional to $e^{iKs}$, where $K$ is the proper angular frequency and $s$ is the proper time. In this way, the epicyclic frequency $\kappa^2=K^2/(u^0)^2$ and the perturbations $\tau_\perp^2$ in direction $x^3$ are respectively calculated as  

\begin{eqnarray}\label{kappa}
\kappa^2 = (\Gamma^1_{22,1}-4\Gamma^2_{21}\Gamma^1_{22}
-4\Gamma^0_{21}\Gamma^1_{20})\Omega^2 +
(2\Gamma^1_{20,1}-4\Gamma^2_{21}\Gamma^1_{02}-4\Gamma^0_{01}\Gamma^1_{20}-\\ \nonumber
-4\Gamma^0_{21}\Gamma^1_{00}-4\Gamma^2_{01}\Gamma^1_{22})\Omega + \Gamma^1_{00,1} - 4 \Gamma^0_{01}\Gamma^1_{00} - 4 \Gamma^2_{01}\Gamma^1_{02} - V^1_{\;\;,1}/(u^0)^2, 
\end{eqnarray}

\begin{equation}\label{tau}
\tau_\perp^2 = \Gamma^3_{00,3}+2\Gamma^3_{02,3}\Omega+\Gamma^3_{22,3}\Omega^2.
\end{equation}

\section{Example 1: General perturbations for $5D$ metric in the Newtonian limit (with no compactification)}\label{ex1}
The main aim now is to compute the gravitational potential in the Newtonian
limit, since galaxies and clusters can be described physically as
Newtonian objects --- corresponding to the approximation in which gravity is
weak. The weak limit is assumed uniquely in the 4-dimensional
spacetime: the deviation $\gamma_{\alpha\beta}$
of the 4-dimensional metric
$g_{\alpha\beta}=\eta_{\alpha\beta}+\gamma_{\alpha\beta}$ is small
 ($\eta_{\alpha\beta}$ denotes the Minkowski metric). Linearized gravity has a gauge freedom given by
$\gamma_{\alpha\beta} \mapsto \gamma_{\alpha\beta} +
\pounds_{\xi}\eta_{\alpha\beta}$, where $\pounds_{\xi}$ denotes the Lie
derivative with respect to the generators $\xi^\alpha$ of a differential
diffeomorphism. To the first order, such transformation represents the
same physical transformation  as $\gamma_{\alpha\beta}$. This gauge freedom is used to simplify the linearized Einstein equation. Solving the equation $\partial^\beta\partial_\beta\xi_\alpha =
-\partial^\beta \overline{\gamma}_{\alpha\beta}$ for $\xi_\alpha$, a gauge transformation that leads to
$\partial^\beta\overline{\gamma}_{\alpha\beta} = 0$  --- similar to the
Lorentz gauge condition --- can be elicited  to obtain the simplified Einstein equation
\bege \label{linear_t}
\mathcal{T}_{\alpha\beta}=-\frac{1}{4}\partial^\mu\partial_\mu
\overline{\gamma}_{\alpha\beta},\enge \noi
\noi and
\bege
\mathfrak{T}_{\alpha\beta} =
\frac{1}{2}\left[\frac{1}{2}(g^{mn}\partial^\mu\partial_\mu
g_{mn})g_{\alpha\beta} - g^{mn}g_{mn,\alpha\beta}\right], \enge
\noi where $\mathfrak{T}_{\alpha\beta}$ refers to terms of the stress tensor dependent of extra terms $g_{ab}$ of the metric. When gravity is weak, the linear approximation to GR should be valid. There exists a global inertial
coordinate system of $\eta_{\alpha\beta}$ such that \bege
T_{\alpha\beta} = \mathcal{T}_{\alpha\beta} +
\mathfrak{T}_{\alpha\beta} \approx \rho t_\alpha t_\beta, \enge
 \bege\label{super_poisson}  -\frac{1}{4}\partial^\mu\partial_\mu
\overline{\gamma}_{\alpha\beta}+\frac{1}{2}\left[\frac{1}{2}(g^{mn}\partial^\mu\partial_\mu
g_{mn})\eta_{\alpha\beta} - g^{mn}g_{mn,\alpha\beta}\right] = \rho
t_{\alpha} t_{\beta},\enge
\noi where $t_{\alpha}$ is the time direction associated with this coordinate
system. This equation can be interpreted as the modified Poisson
equation considering a universe with more than $3+1$ dimensions.

Define $\overline{\gamma}_{\alpha\beta} \equiv -4\phi$, where
$\phi=\phi(\vec{x})$ is a 3-space scalar field. Furthermore consider a line element $\mathrm{d}s_n^2=\sum_{i=1}^{n}
e^{\psi_i}\mathrm{d}z_i^2$, where
$\mathrm{d}s^2_n$ is the world line for the extra sector, $z_i$ denotes
the extra coordinates and $\psi_i=\psi_i(\vec{x})$ are
potentials associated to extra dimensions.


If one asserts, as a first approximation the sigma
model $g^{\mu\nu}(\sigma_{,\mu}\sigma^{-1})_{,\nu}=0$ for the extra
part, where $\sigma$ denotes the diagonal matrix representing the metric
associated to the system, we have	
\bege\label{laplace_extra}\partial^\mu\partial_\mu g_{ab} = 0,\enge
\noi yielding the following equation
\bege\label{super_poisson_2}  -\frac{1}{4}\partial^\mu\partial_\mu
\overline{\gamma}_{\alpha\beta}-\frac{1}{2}g^{mn}g_{mn,\alpha\beta}
= \rho t_{\alpha} t_{\beta},\enge \noi or in other words
\bege \label{poisson}\nabla^2 \phi = \rho. \enge
\noi It means that our visible matter density profile is provided
uniquely by the 4-dimensional field. In the $5D$ Newtonian limit one can write the following line element

\begin{equation}
\mathrm{d}s^2 =
-(1-2\phi)\mathrm{d}t^2 + \mathrm{d}\vec{x}.\mathrm{d}\vec{x} +
e^{-\psi}\mathrm{d}y_1^2,
\end{equation}
where $\mathrm{d}\vec{x}\cdot\mathrm{d}\vec{x}$ is the
$3$-dimensional line element and $y_ 1$ is the extradimension. In cylindrical coordinates
the 3D line element will be $\mathrm{d}\vec{x}\cdot\mathrm{d}\vec{x}=dr^2+r^2d\varphi^2+dz^2$.
 
To find a form for those functions $\phi$ and $\psi$, from (\ref{laplace_extra}) and
(\ref{poisson}) it yields
\begin{eqnarray}\label{laplace_equation}
\nabla^2 \psi - \nabla \psi \cdot \nabla \psi = 0\\
\label{phi}
\nabla^2 \phi = \rho.
\end{eqnarray}
\noindent Non-linear terms do not appear, since the $\sigma$
matrix is diagonal. In particular, Eq. (\ref{laplace_equation}) can
be rewritten as
\begin{equation}\label{laplace_2}
\nabla^2 \chi=0,
\end{equation}
\noindent where the identification $
\chi = e^{-\psi}
$ is accomplished.

\noi Simple solutions for those functions are, for example,

\begin{equation}
\phi =  - \frac{m}{\sqrt{r^2 + (\bar{z}+a)^2}},
\end{equation}
and
\begin{equation}\label{chi}
\chi = \frac{2m}{r},
\end{equation}
\noi where the coordinate $\bar{z}=|z|+c$ introduces a cut method to generate a disk solution, where $c>0$ is the cut parameter. Here, $a$ is a general constant, and the above solutions are given for a particle of mass $m$ in the position $z\rightarrow 0$. The solution (\ref{chi}) gives

\begin{equation}
\psi = - \ln \chi.
\end{equation}

\noi The epicyclic $\kappa^2$ frequency is calculated from Eq. (\ref{kappa}) as

\begin{equation}\label{eq_kappa_ex1}
\kappa^2 (r)= \left[3-\frac{1}{2}N_{y1}^2r^2e^\psi
\left(\frac{d^2\psi}{dr^2}+\left(\frac{d\psi}{dr}\right)^2\right)\right]\Omega^2 -\frac{4}{1-2\phi}\left(\frac{d \phi}{dr}\right)^2 - \frac{d^2 \phi}{dr^2} + \frac{1}{2}e^\psi(1-2\phi)N_{y1}^2
\left[\frac{d^2\psi}{dr^2}+\left(\frac{d\psi}{dr}\right)^2\right],
\end{equation}

\noi where the squared angular velocity $\Omega^2$ is calculated from $\Omega^2 (u^0)^2=F(r)=(-g_{22}/g_{00})(\dot{\varphi}^2/\dot{t}^2)$ (see e.g. \cite{coimbrasss1}), from (\ref{ut}) and (\ref{constant}) as

\begin{equation}
\Omega^2 = \frac{F(r)(1-2\phi)}{1+F(r)r^2},
\end{equation}

\noi with 
\begin{equation}
F(r)= \frac{r}{1-2\phi} \left[\frac{e^\psi \partial_r \phi + (e^\psi \partial_r \psi)/2 - \phi e^\psi \partial_r \psi - \partial_r \phi/N_{y1}^2}{(e^\psi r\partial_r \psi)/2 - e^\psi + 1/N_{y1}^2} \right].
\end{equation}

\noi The orthogonal perturbation $\tau_\perp^2$ is calculated simply from (\ref{tau}) as

\begin{equation}\label{eq_tau_ex1}
\tau_\perp^2 = - \frac{\partial^2 \phi}{\partial z^2},
\end{equation}

\noi where in this last case $\tau_\perp^2$ is plotted for all $z>0$ and $r>0$. Fig. \ref{fig:kappa_newtonian}(a) and Fig. \ref{fig:kappa_newtonian}(c) show respectively the curves for $\kappa^2$ and $\tau_\perp^2$ for some values of $N_{y1}$. 

\section{Example 2: $5D$ Randall-Sundrum (compactified-like)} \label{ex2}

\subsection{Pure Randall-Sundrum}\label{ex21}

The Randall-Sundrum (RS) metric is in general expressed as
\begin{equation}\label{rs_metric}
ds^2 = e^{-2k|y|}g_{\mu\nu}dx^{\mu}dx^{\nu} + dy^2,
\end{equation}
\noindent
where $k^2 = 3/(2\ell^2)$, and the term $e^{-2k|y|}$ is called \emph{the warp factor} \cite{braneworld,braneworld0}, which
reflects the confinement role of a extradimensional anti-de Sitter bulk constant $\Lambda$ that prevents gravity from leaking into the extra dimension at low energies. The term $|y|$ provides the $\mathbb{Z}_2$ symmetry of the 3-brane at $y=0$ and RS metric can be regarded as an alternative to compactification.

\begin{figure}
  \centering
  \includegraphics[angle=90,width=0.9\textwidth]{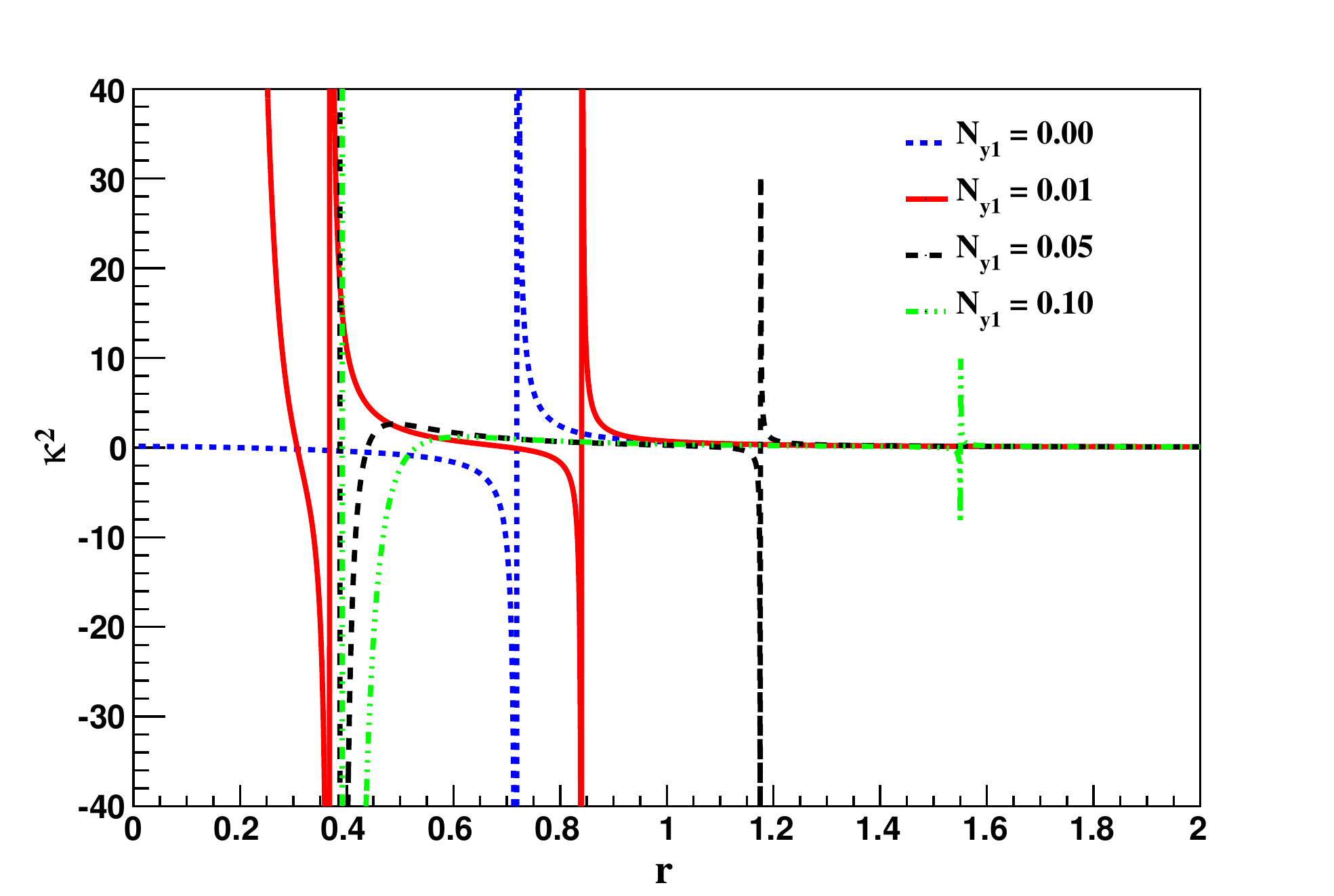}
  \caption{The plots show $\kappa^{2}$ as a function of the
    $r$ for the general Randall-Sundrum $5D$ metric with Weyl coordinates. The perturbed geodesics of the pure RS with Weyl coordinates is recovered when $N_{y1}=0$.}
  \label{fig:kappa_RS}
\end{figure}

Here we will assume that $g_{\mu\nu}$ is the Weyl axisymmetric metric described in (\ref{weyl}). The perturbed geodesics of this pure RS with Weyl coordinates is represented in Fig. \ref{fig:kappa_RS} for $N_{y1}=0$, that is a particular case of the second example below.

\subsection{General Randall-Sundrum}\label{ex22}

Another possibility is that one can assume a general Randall-Sundrum metric with Weyl-Lewis-Papapetrou coordinates written as

\bege
ds^2= e^{-2k|y|}g_{\mu\nu}dx^{\mu}dx^{\nu} + e^{-\psi}dy^2,
\enge

\noi with the assumption discussed in Section \ref{stabilitynd}, i.e., that $g_{ym}\dot{x}^m = N_y$, with $g_{yy}=e^{-\psi}$. When $N_y=0$, it is recovered the original pure RS metric (\ref{rs_metric}). With $g_{\mu\nu}$ given by (\ref{weyl}) with solutions

\begin{equation}
\lambda = \xi =  - \frac{m}{\sqrt{r^2 + (\bar{z}+a)^2}},
\end{equation}
and
\begin{equation}
\psi = \frac{2m}{r}.
\end{equation}

\noi Here we have introduced the same cut method as before to generate a disk solution, such that $\bar{z}= |z|+c$, where $c>0$ is the cut parameter. The epicyclic $\kappa^2$ frequency calculated from Eq. (\ref{kappa}) as

\begin{eqnarray}\label{eq_kappa_ex22}
\begin{split}
\kappa^2 (r)&= \left[3 - 4r\left(\frac{d
      \lambda}{dr}\right)+2r^2\left(\frac{d
      \lambda}{dr}\right)^2+r^2\frac{d^2
    \lambda}{dr^2}\right]e^{-2\lambda}\Omega^2 -\\
& \left[2\left(\frac{d \lambda}{dr}\right)^2-\frac{d^2 \lambda}{dr^2}
 \right]e^{2\lambda} - e^{2\lambda}V^1_{\;\;,1}+e^{-2\lambda}r^2V^1_{\;\;,1}\Omega^2
\end{split}
\end{eqnarray}

\noi The squared angular velocity $\Omega$ is
\begin{equation}
\Omega^2 = \frac{e^{2\lambda}F(r)}{1+r^2e^{-2\lambda}F(r)},
\end{equation}
and
\begin{equation}\nonumber
F(r)=\frac{H(r)}{W(r)},
\end{equation}
with,
\begin{equation}
H(r)=re^{-4\lambda}[2N_{y}e^{\psi}(\partial_r\lambda)-N_{y}e^{\psi}(\partial_r\psi)-2(\partial_r\lambda)],
\end{equation}
\begin{eqnarray}
W(r)=2e^{-4\lambda}-2rN_{y}e^{-4\lambda}e^{\psi}(\partial_r\lambda)+N_{y}re^{-4\lambda}e^{\psi}(\partial_r\psi)+\\
+2re^{-4\lambda}(\partial_r\lambda)-2N_{y}e^{-4\lambda}-4rN_{y}(\partial_r\lambda)e^{-4\lambda}-4r(\partial_r\lambda)e^{-4\lambda},
\end{eqnarray}
and
\begin{equation}
V^1_{\;\;\;,1} = -\frac{1}{2}N_{y}^{2}e^{-\lambda}e^{\psi}(\partial_r^{2}\psi)-\frac{1}{2}N_{y}^{2}e^{-\lambda}(\partial_r\psi)^2-2N_{y}^{2}e^{\psi}(\partial_r\lambda)(\partial_r\psi).
\end{equation}

\noi The orthogonal perturbations are calculated from (\ref{tau}) as
\begin{equation}\label{eq_tau_ex22}
\tau_\perp^2 = [2\lambda_{,z}-2\xi_{,z}+\xi_{,zz}+2(\xi_{,z})^2]e^{4\xi - 2\lambda}+[\xi_{,zz}-2\lambda_{,z}\xi_{,z}]e^{-2\lambda}\Omega^2.
\end{equation}

Fig. \ref{fig:kappa_RS} and Fig. \ref{fig:tau_RS} show respectively the curves for $\kappa^2$ and $\tau_\perp^2$ for some values of $N_{y1}$. The system is stable if $\kappa^2>0$ and $\tau_\perp >0$. When $N_{y1} \rightarrow 0$, the curves are stable both for the radial and orthogonal perturbations for $r > 0.7$. Larger the values of $N_{y1}$, more instabilities are present.

\section{Example 3: $6D$ Newtonian limit for an axial configuration}\label{ex3}

The metric for the Newtonian limit presented in Sec. \ref{ex1} can be expanded for the case of a $6D$ configuration as

\begin{equation}
\mathrm{d}s^2 =
-(1-2\phi)\mathrm{d}t^2 + \mathrm{d}\vec{x}.\mathrm{d}\vec{x} +
e^{-\psi}\mathrm{d}y_1^2 + e^{\psi}\mathrm{d}y_2^2
\end{equation}
where $\mathrm{d}\vec{x}\cdot\mathrm{d}\vec{x}$ is the
$3$-dimensional line element and $y_ 1$ and $y_2$ are the extradimensional coordinates. In cylindrical coordinates
the 3D line element will be $\mathrm{d}\vec{x}\cdot\mathrm{d}\vec{x}=dr^2+r^2d\varphi^2+dz^2$.

From (\ref{laplace_extra}) and (\ref{poisson}), simple solutions for those functions are (where, as stated before, $\bar{z}$ generates the disk solution)

\begin{equation}
\phi =  - \frac{m}{\sqrt{r^2 + (\bar{z}+a)^2}},
\end{equation}
and
\begin{equation}
\chi = \frac{2m}{r},
\end{equation}
where $a$ is a constant, and both solutions are given for a particle of mass $m$ in the position $z\rightarrow 0$. In this case,

\begin{equation}
\psi = - \ln \chi.
\end{equation}

\noi The epicyclic $\kappa^2$ frequency is calculated from Eq. (\ref{kappa}) as

\begin{equation}\label{eq_kappa_ex3}
\kappa^2 (r)=  (3 + r^2V^1_{\;\;,1})\Omega^2 -\frac{4}{1-2\phi}\left(\frac{d \phi}{dr}\right)^2 - \frac{d^2 \phi}{dr^2} - V^1_{\;\;,1}(1-2\phi),
\end{equation}

\noi where $V^1_{\;\;,1}$ and the squared angular velocity $\Omega^2$ are calculated as

\begin{equation}
V^1_{\;\;,1} = -\frac{1}{2}N_{y1}^{2}e^{\psi}(\partial_r\psi)^2-\frac{1}{2}N_{y2}^{2}e^{-\psi}(\partial_r\psi)^2-\frac{1}{2}N_{y1}^2 e^{\psi}(\partial^2_r\psi)+\frac{1}{2}N_{y2}^2e^{-\psi}(\partial^2_r\psi).
\end{equation}

\begin{equation}
\Omega^2= \frac{(1-2\phi) F(r)}{1+r^2 F(r)},
\end{equation}

\begin{equation}
H(r)=N_{y1}^2e^\psi[(\partial_r\psi)-2\phi(\partial_r\psi)-2(\partial_r\phi)]-N_{y2}^2
e^{-\psi}[(\partial_r\psi)-2\phi(\partial_r\psi)+2(\partial_r
\phi)]-2(\partial_r \phi)
\end{equation}

\noi with

\begin{equation}
F(r) = \frac{r}{1-2\phi}\left[\frac{H(r)}{N_{y1}^2 e^\psi[r(\partial_r \psi)+2]-N_{y2}^2 e^{-\psi}[r(\partial_r \psi)-2]+2}\right].
\end{equation}
\noi The orthogonal perturbation $\tau_\perp^2$ is calculated simply as

\begin{equation}\label{eq_tau_ex3}
\tau_\perp^2 = - \frac{\partial^2 \phi}{\partial z^2},
\end{equation}

\noi where in this last case $\tau_\perp^2$ is plotted for all $z>0$ and $r>0$. Fig. \ref{fig:kappa_newtonian}(b) and Fig. \ref{fig:kappa_newtonian}(c) show respectively the curves for $\kappa^2$ and $\tau_\perp^2$ for some values of $N_{y1}$ and $N_{y2}$. Note that in this case, there are no differences between $5D$ and $6D$ perpendicular perturbations.

\section{Example 4: $6D$ Randall-Sundrum}\label{ex4}

We propose a $6D$ general Randall-Sundrum (RS) metric written as
\begin{equation}\label{rs_metric6d}
ds^2 = e^{-2k|y|}g_{\mu\nu}dx^{\mu}dx^{\nu} + e^\psi dy_1^2 +e^{-\psi} dy_2^2.
\end{equation}

\noi Here it is assumed again that $g_{\mu\nu}$ is the Weyl axisymmetric metric described in (\ref{weyl}), with the assumption discussed in Section \ref{stabilitynd}, i.e., that $g_{55}\dot{x}^5 = N_{y1}$ and $g_{66}\dot{x}^6 = N_{y2}$, with $g_{55}=e^\psi$ and $g_{66}=e^{-\psi}$. Ther term $|y|$ in the warp factor $e^{-2k|y|}$ is $|y|=\sqrt{y_1^2+y_2^2}$. The solutions are

\begin{equation}
\lambda = \xi =  - \frac{m}{\sqrt{r^2 + (\bar{z}+a)^2}},
\end{equation}
and
\begin{equation}
\psi = \frac{2m}{r}.
\end{equation}

\noi The epicyclic $\kappa^2$ frequency calculated from Eq. (\ref{kappa}) is

\begin{eqnarray}\label{eq_kappa_ex4}
\kappa^2 (r)&= \left[3 - 4r\left(\frac{d
      \lambda}{dr}\right)+2r^2\left(\frac{d
      \lambda}{dr}\right)^2+r^2\frac{d^2
    \lambda}{dr^2}\right]e^{-2\lambda}\Omega^2 -\\
& \left[2\left(\frac{d \lambda}{dr}\right)^2-\frac{d^2 \lambda}{dr^2}
 \right]e^{2\lambda}
- e^{2\lambda}V^1_{\;\;,1}+e^{-2\lambda}r^2V^1_{\;\;,1}\Omega^2,
\end{eqnarray}

\noi with

\begin{equation}
V^1_{\;\;,1} = -\frac{1}{2}N_{y1}^{2}e^{-\psi}(\partial_r\psi)^2-\frac{1}{2}N_{y2}^{2}e^{\psi}(\partial_r\psi)^2+\frac{1}{2}N_{y1}^2 e^{-\psi}(\partial^2_r\psi)-\frac{1}{2}N_{y2}^2e^{\psi}(\partial^2_r\psi).
\end{equation}

\noi The squared angular velocity $\Omega$ is
\begin{equation}
\Omega^2 = \frac{e^{2\lambda}F(r)}{1+r^2e^{-2\lambda}F(r)},
\end{equation}
and
\begin{equation}\nonumber
F(r)=\frac{H(r)}{W(r)},
\end{equation}
with,
\begin{equation}
H(r)=re^{-4\lambda}[(2N_{y1}e^{-\psi}+2N_{y2}e^{\psi})(\partial_r\lambda)-(N_{y1}e^{\psi}-N_{y2}e^{-\psi})(\partial_r\psi)-2(\partial_r\lambda)],
\end{equation}
\begin{eqnarray}
\begin{split}
W(r)&=2e^{-4\lambda}-2r(N_{y1}e^{\psi}+N_{y2}e^{-\psi})(e^{-4\lambda}\partial_r\lambda)+(N_{y1}e^{\psi}-N_{y2}e^{-\psi})re^{-4\lambda}(\partial_r\psi)+\\
&+2re^{-4\lambda}(\partial_r\lambda)-2(N_{y1}+N_{y2})e^{-4\lambda}-4r(N_{y1}+N_{y2})(\partial_r\lambda)e^{-4\lambda}- \\
&-4r(\partial_r\lambda)e^{-4\lambda}
\end{split}
\end{eqnarray}

\noi The orthogonal perturbations are calculated from (\ref{tau}) as
\begin{equation}\label{eq_tau_ex4}
\tau_\perp^2 = [2\lambda_{,z}-2\xi_{,z}+\xi_{,zz}+2(\xi_{,z})^2]e^{4\xi - 2\lambda}+[\xi_{,zz}-2\lambda_{,z}\xi_{,z}]e^{-2\lambda}\Omega^2.
\end{equation}

\noi Figs. \ref{fig:kappa_RS6} and \ref{fig:tau_RS6} show the behavior of $\kappa^2$ and $\tau_\perp^2$ for some values of $N_{y1}$ and $N_{y2}$.

\section{Discussion and concluding remarks}

In the present work perturbative terms were calculated explicitly in particle motion due to the presence of extradimensions (compactified or not) in an axially symmetric configuration. We showed that extradimensions add terms to the original perturbative equation of classical particle geodesics for any geometry. First of all, we calculated the equation of motion (\ref{motion}) and, with the transformation $x^\alpha \rightarrow x^\alpha + \Delta^\alpha$, it is possible to find the equation (\ref{perturbativend}) for perturbations in particle geodesic motion with the presence of extradimensions. The term $V^\alpha_{,\gamma}\Delta^\gamma$ arises and we test the epyciclic radial frequency $\kappa^2$ (\ref{kappa}) and axial oscillations $\tau_\perp^2$ (\ref{tau})  for Weyl metric in cilindric coordinates for $5D$ and $6D$ configurations. We showed that when $\kappa^2>0$ and $\tau_\perp^2>0$ the system is stable both in radial as in axial directions.

Five metrics were used to calculate $\kappa^2$ and $\tau_\perp^2$, namely (i) the Newtonian limit of a general proposed $5D$ [Sec. \ref{ex1}, eqs. (\ref{eq_kappa_ex1}) and (\ref{eq_tau_ex1})] and $6D$ [Sec. \ref{ex3}, eqs. (\ref{eq_kappa_ex3}) and (\ref{eq_tau_ex3})] axisymmetric spacetimes; (ii) a simple Randall-Sundrum $5D$ spacetime [Sec. \ref{ex21}]; (iii) general $5D$ [Sec. \ref{ex22}, eqs. (\ref{eq_kappa_ex22}) and (\ref{eq_tau_ex22})]  and $6D$ [Sec. \ref{ex4}, eqs. (\ref{eq_kappa_ex4}) and (\ref{eq_tau_ex4})] Randall-Sundrum spacetimes.

In all cases, the solutions for the metric potentials that are used to compute the oscillations have a reflexion symmetry in the axial coordinate to create an infinite thin disk of matter. This occurs because we have the coordinate $\bar{z}=|z|+c$ in the solutions, where $c$ is the disk cut parameter. The matter in the disk comes from the discontinuity in the stress tensor when $z\rightarrow 0$, since $\partial_z |z|=2 \vartheta(z)-1$ and $\partial_{zz} |z|=2\delta(z)$, where $\vartheta(z)$ and $\delta(z)$ are, respectively, the Heaviside function and the Dirac distribution. Therefore the Einstein field equations will be separated in two different pieces: one valid for $z\not =0$ (the usual Einstein equations), and other involving distributions with an associated energy-momentum tensor. Due to the discontinuous behavior of the derivatives of the metric tensor across the disk, the Riemann curvature tensor contains Dirac delta functions. The energy-momentum tensor can be obtained by the distributional approach due to Papapetrou and Hamouni \cite{papapetrou}, Lichnerowicz \cite{lichnerowicz}, and Taub \cite{taub}. It can be written as ${T^\alpha}_\beta = [{T^\alpha}_\beta] \ \delta(z)$, where $\delta$ is the Dirac function with support on the disk and $[{T^\alpha}_\beta]$ is the distributional energy-momentum tensor, which yield the volume energy density and the principal stresses. The disk at $z=0$ divides the space-time into two halves. The normal to the disk can be described by the co-vector $n_a=\partial z/\partial x^a=(0,0,0,1)$. Above the disk near $z=0$, we can expand the metric as
\begin{equation}
g_{\alpha\beta}=g^0_{\alpha\beta} + z\frac{\partial g_{\alpha\beta}^+}{\partial z}\vert_{z=0}+z^2\frac{\partial^2 g_{\alpha\beta}^+}{\partial z^2}\vert_{z=0}+...,
\end{equation} and below $z=0$,
\begin{equation}
g_{\alpha\beta}=g^0_{\alpha\beta} + z\frac{\partial g_{\alpha\beta}^-}{\partial z}\vert_{z=0}+z^2\frac{\partial^2 g_{\alpha\beta}^-}{\partial z^2}\vert_{z=0}+....
\end{equation} The quantity $g^0_{\alpha\beta}$ means the value of $g_{\alpha\beta}$ at $z=0$. The discontinuities in the first derivatives of the metric tensor can be cast as $b_{\alpha\beta} \ = \ g_{\alpha\beta,z}|_{_{z=0^+}} \ - \ g_{\alpha\beta,z}|_{_{z = 0^-}}$ in such manner that $\left[ {\Gamma^\alpha}_{\beta\gamma} \right] = \frac{1}{2}({b^\alpha}_\gamma{\delta^z}_{\beta} + {b^\alpha}_\beta{\delta^z}_{\gamma} - g^{\alpha z}b_{\beta\gamma})$ where $\left[ {\Gamma^\alpha}_{\beta\gamma} \right] \equiv {\Gamma^{+\alpha}}_{\beta\gamma} - {\Gamma^{-\alpha}}_{\beta\gamma}$ at $z=0$. In this way, we can identify the distributional energy-momentum tensor on the disk through Einstein equations as $[{R^\alpha}_{\beta}] - \frac{1}{2}{\delta^\alpha}_b[R]=8\pi [{T^\alpha}_\beta]$. Then the distributional energy-momentum tensor is given by $[{T^\alpha}_{\beta}]=\frac{1}{16\pi}\{b^{\alpha z}{\delta^z}_{\beta}-b^{zz}\delta^{\alpha}_{\beta}+g^{\alpha z}{b^{z}}_\beta-g^{zz}{b^{\alpha}}_\beta+ {b^{\gamma}}_\gamma(g^{zz}{\delta^{\alpha}}_{\beta}-g^{\alpha z}{\delta^{z}}_{\beta})\}$. The energy density $\epsilon$ and pressures $p_\alpha$ in the disk are calculated for the developed axisymmetric configurations as $\epsilon = -[{T^t}_t]$, $p_\varphi = [{T^\varphi}_{\varphi}]$, $p_r = [{T^r}_r] = 0$, $p_z = [{T^z}_z] = 0$. The energy conditions are always satisfied for the four examples bellow. Specifically in examples 2 and 4 it is satisfied when the extra coordinate $y\rightarrow 0$, i.e., when we analyze the stability in the $3D$ disk.

Concerning the method to calculate the oscillations and the stability of the system, in general, it is verified that extra dimensions contribute to destabilize the disk, but stability is verified for some cases. In what follows, we present a summary of the most important points to be discussed from the mentioned examples. 

\begin{figure}
  \centering
  {\includegraphics[angle=90,width=0.6\textwidth]{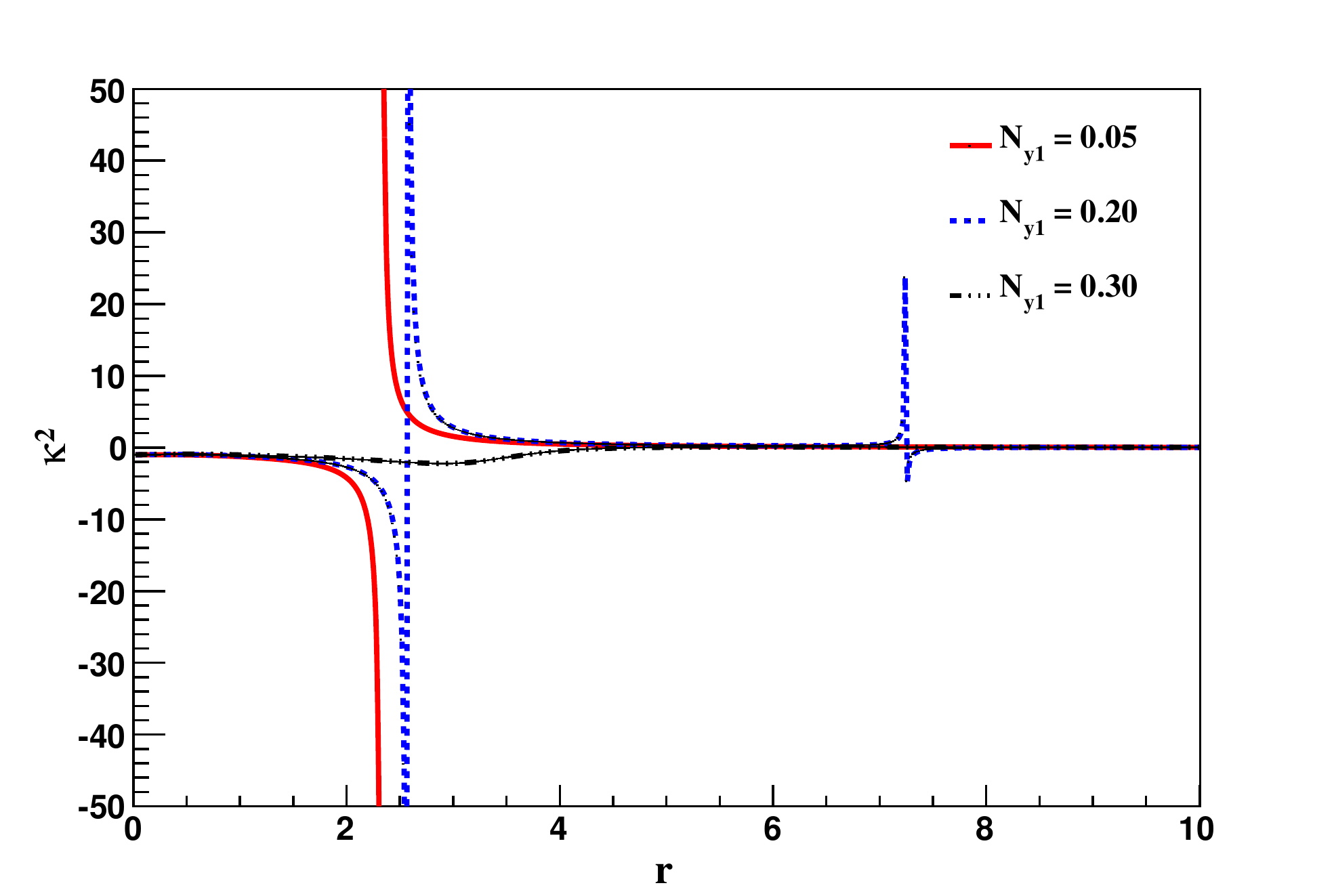}}\\(a) Stability for the Newtonian limit (axial symmetry) in $5D$.\\
  {\includegraphics[angle=90,width=0.6\textwidth]{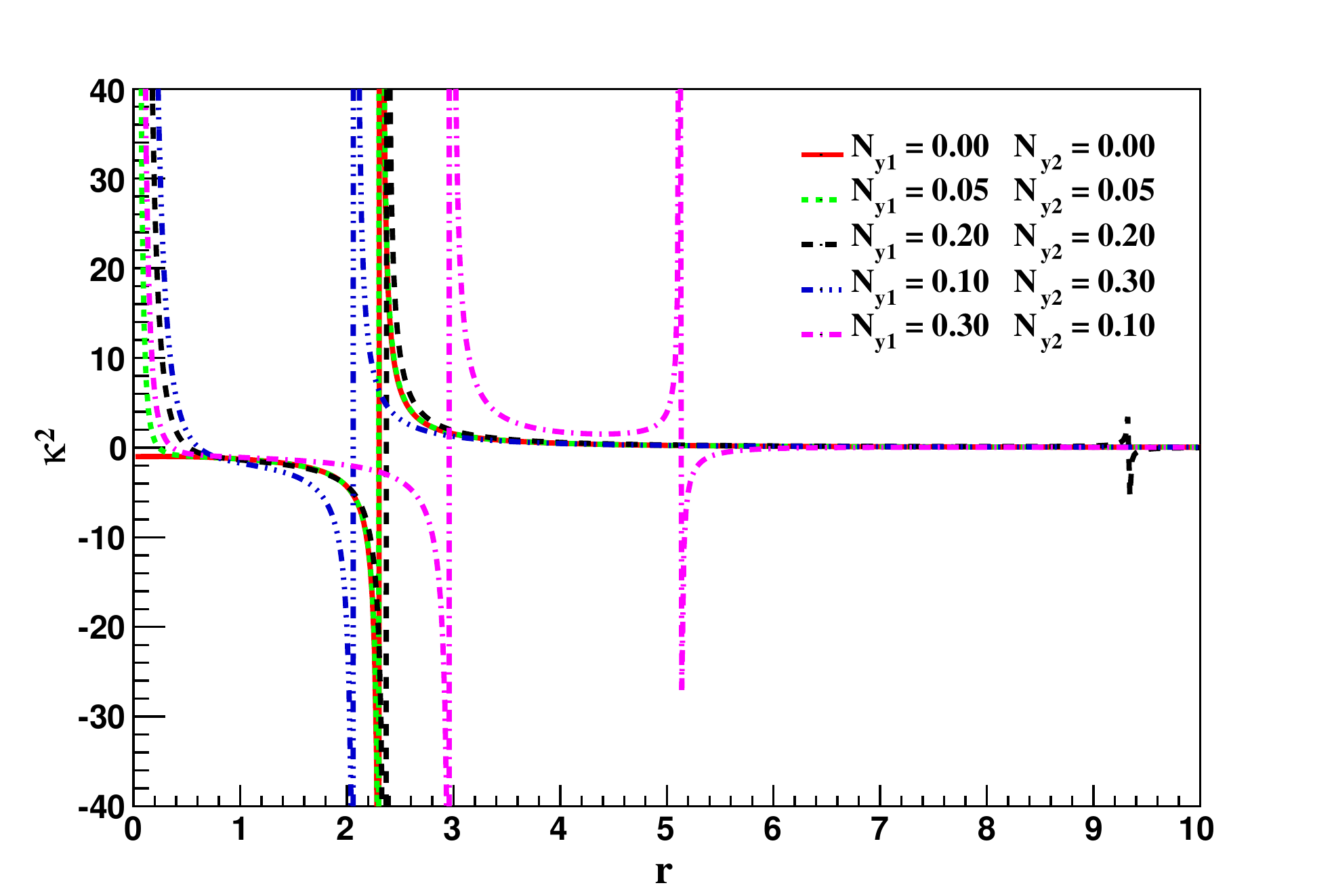}}\\(b)Stability for the Newtonian limit (axial symmetry) in $6D$.\\
  {\includegraphics[angle=0,width=0.6\textwidth]{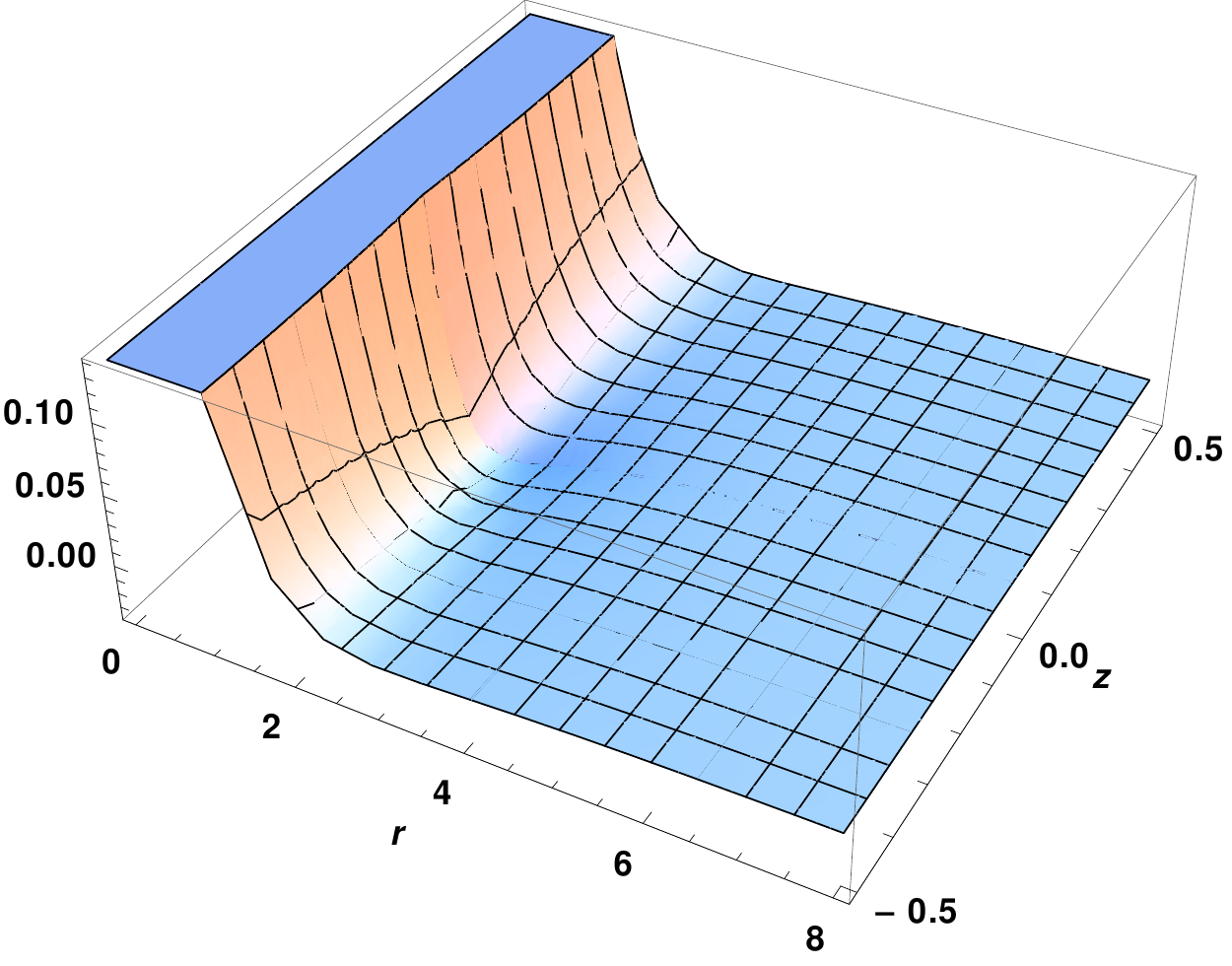}}\\(c) $\tau_\perp^{2}$ as a function of $r$ and $z$.
  \caption{(a) The plots show $\kappa^{2}$ as a function of the
    $r$ coordinate. The system is stable if $\kappa^2>0$. When $N_{y1}
    \rightarrow 0$, the curves are stable. Larger the values of
    $N_{y1}$, more instabilities are present. (b) The plots show $\kappa^{2}$ as a function of the
    $r$ coordinate. The system is stable if $\kappa^2>0$. (c) The plot shows $\tau_\perp^{2}$ as a function of
    $r$ and $z$ for axisymmetric coordinates in the Newtonian limit in
    $5D$ and $6D$ to $a=0$ and cut parameter is $c=1$ (satisfying energy conditions). The system is stable when $\tau_\perp^2 >0$. }
  \label{fig:kappa_newtonian}
\end{figure}

Example 1 - Newtonian limit of a general proposed $5D$ axial symmetry (Sec. \ref{ex1}): Fig. \ref{fig:kappa_newtonian}(a) and Fig. \ref{fig:kappa_newtonian}(c) show respectively the curves for $\kappa^2$ and $\tau_\perp^2$ for some values of $N_{y1}$ (integration constant thanks to extradimension) in the case where the $5D$ axisymmetric system is in the Newtonian limit. The system is stable if $\kappa^2>0$ and $\tau_\perp^2 >0$. When $N_{y1} \rightarrow 0$, the curves are stable both for the radial and orthogonal perturbations. In this case, it is recovered the $4D$ expected stability. Larger the values of $N_{y1}$, more instabilities are present. It is possible to see this indeed in expression (54) since $N_{y1}$ is associated to negative terms of $\kappa^2$. Nevertheless, there are indeed some values of $N_{y1}$ where the perturbed axial system presents stability inside a region of the axial system between $0 < r < r_s$. For example, Fig. \ref{fig:kappa_newtonian}(a) shows other two cases. When $N_{y1}=0.20$, $r_s \approx 7$. When $N_{y1} > 0.30$ the results show that $\kappa^2 <0$ for all values of $r$, and therefore the system for such case is unstable. The perpendicular perturbations of Fig. \ref{fig:kappa_newtonian}(c) show that the system is stable for all $r$ range. In fact, from $5D$ axisymmetric Newtonian limit, it is possible to conclude that {\bf one extradimension indeed carries instabilities for the system in the radial direction and does not carry instabilities in the axial direction}. 

\begin{figure}
  \centering
  \includegraphics[width=0.7\textwidth]{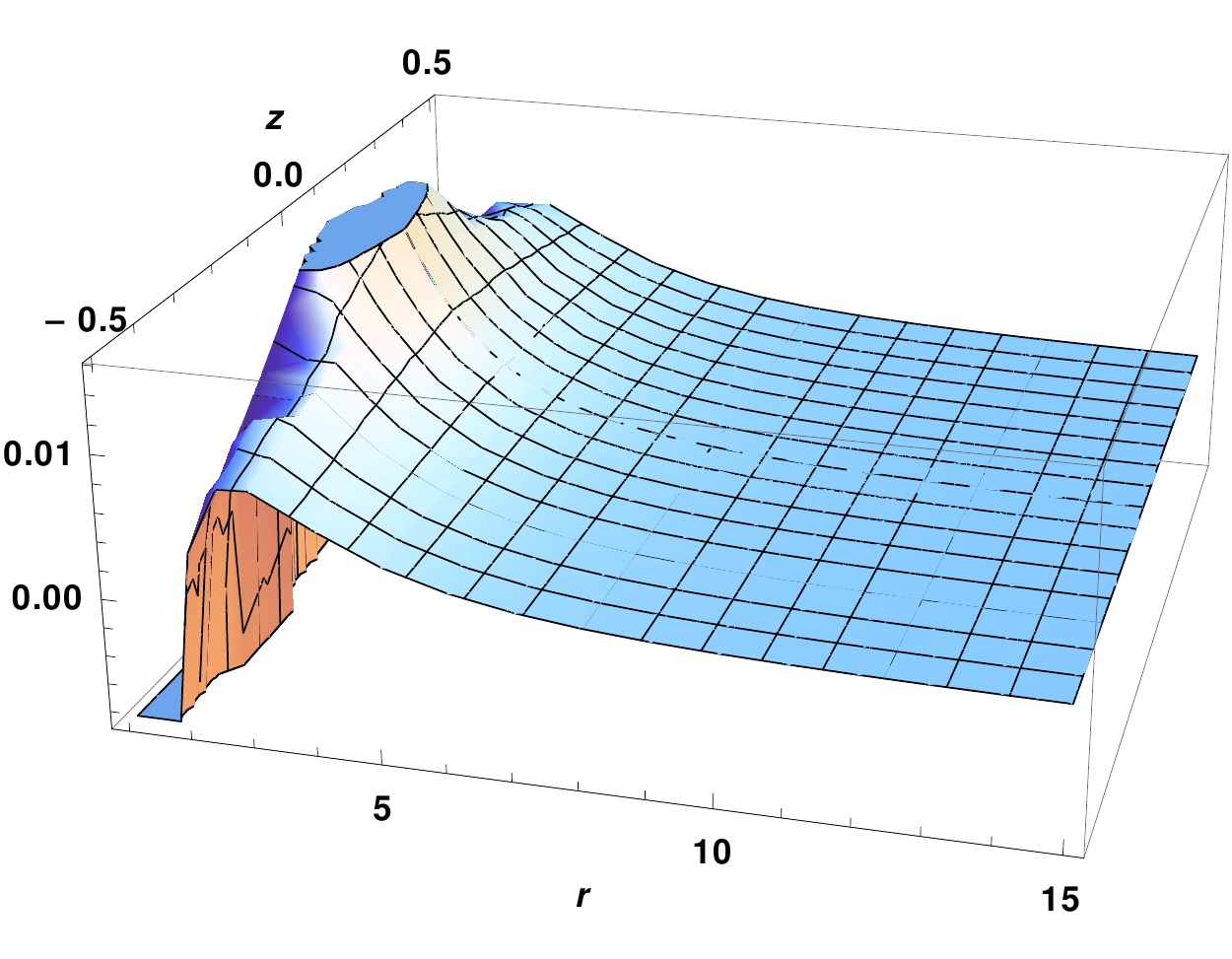}
  \caption{The plot shows $\tau_\perp^{2}$ as a function of
    $r$ and $z$ for the general Randall-Sundrum $5D$ metric with Weyl
    coordinates. Here it is assumed $N_y=0.2$, $a=0$ and and cut parameter is $c=1$ (satisfying energy conditions).}
  \label{fig:tau_RS}
\end{figure}

Example 2 - general $5D$ Randall-Sundrum with axial symmetry
(Sec. \ref{ex2}): Fig. \ref{fig:kappa_RS} and Fig. \ref{fig:tau_RS} show
respectively the curves for $\kappa^2$ and $\tau_\perp^2$ for some
values of $N_{y1}$ in the case where the $5D$ axisymmetric system is
Randall-Sundrum. The system is stable if $\kappa^2>0$ and $\tau_\perp^2
>0$. When $N_{y1} \rightarrow 0$, the curves are stable both for the
radial and orthogonal perturbations if $r>0.7$ (this is the pure RS
system). Larger the values of $N_{y1}$, more instabilities are
present. It is possible to see this indeed in expression (67) since $N_{y1}$ is associated to negative terms of $\kappa^2$. Nevertheless, there are indeed some values of $N_{y1}$ where
the axial system presents stability. For example, for values $N_{y1}
\leq 0.10$, the system is stable inside a region of the axial system,
between $r_{s1} < r < r_{s2}$. In the case of $N_{y1}=0.01$, $r_{s1}
\approx 0.5$ and $r_{s2}
\approx 0.8$. When $N_{y1} > 0.10$ the results show that $\kappa^2 <0$
for all values of $r$, and therefore the system for such case is
unstable. The perpendicular perturbations of Fig. \ref{fig:tau_RS} are
plotted for $N_{y1} = 0.20$ and indicates that in the disk plane
($z\rightarrow 0$) the system is stable only in the radial $r>2$
range. When $z$ acquires greater values (both negative or positive), this range of stable radial regions is also greater. From $5D$ axisymmetric RS, it is possible to conclude that {\bf one extradimension carries instabilities for the system in the radial direction and does not carry instabilities in the axial direction}. Also, the {\bf $5D$ RS case presents \bf more instabilities than the $5D$ Newtonian axisymmetric system}.

Example 3 - Newtonian limit of a general proposed $6D$ axial symmetry
(Sec. \ref{ex3}): Fig. \ref{fig:kappa_newtonian}(b) and
Fig. \ref{fig:kappa_newtonian}(c) show respectively the curves for
$\kappa^2$ and $\tau_\perp^2$ for some values of $N_{y1}$ and $N_{y2}$
in the case where the $6D$ axisymmetric system is in the Newtonian
limit. The system is stable if $\kappa^2>0$ and $\tau_\perp^2 >0$. When
$N_{y1} \rightarrow 0$ and $N_{y2} \rightarrow 0$, the curves are stable
both for the radial and orthogonal perturbations (if nevertheless $r >
2.2$). The results show that any values of $N_{y1}$ and $N_{y2}$ give
stable results. In fact, there are a small range of $r$'s (between 0 and
$\sim 2.1$) that represents an unstable region. This is a central region
of a Miyamoto-Nagai gravitational potential $\phi = - \frac{m}{\sqrt{r^2
    + (\bar{z}+a)^2}}$ (with $\bar{z}= |z| + c$), and such instabilities are indeed expected \cite{miyamoto}. The perpendicular perturbations of Fig. \ref{fig:kappa_newtonian}(c) indicates that the system is always stable. From $6D$ axisymmetric Newtonian limit, it is possible to conclude that two extradimensions does not carry instabilities for the system in the radial direction and also does not carry instabilities in the axial direction. There are only local instabilities represented in Fig. \ref{fig:kappa_newtonian}(b) e.g. by the peaks around $r \approx 9$ ($N_{y1}=0.20$ and $N_{y2}=0.20$) and around $r \approx 5$ ($N_{y1}=0.30$ and $N_{y2}=0.10$). Also it important to highlight that $\tau_\perp^2$ is the same both for $5D$ and $6D$ Newtonian limit cases.

\begin{figure}
  \centering
  \includegraphics[angle=90,width=0.9\textwidth]{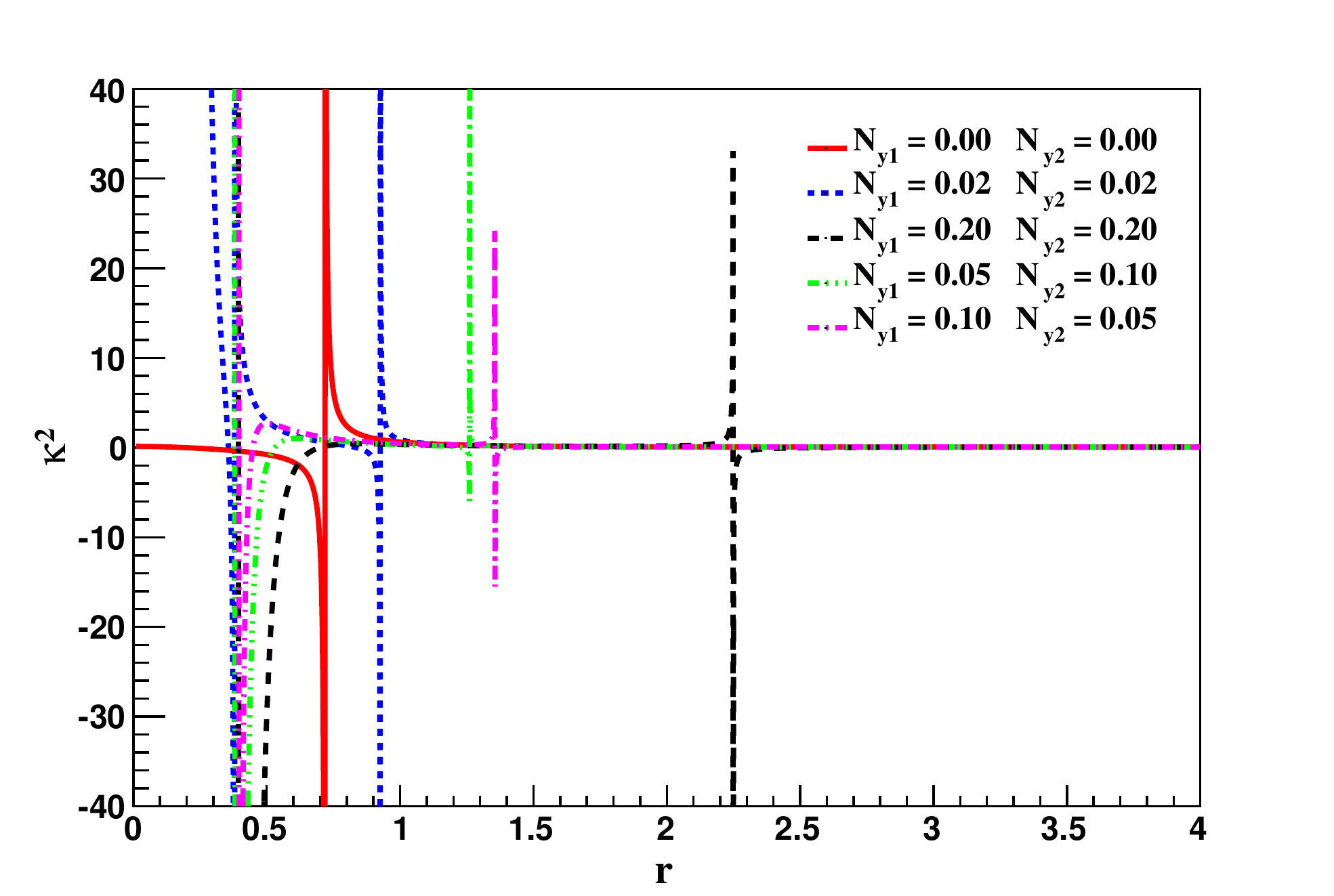}
  \caption{The plots show $\kappa^{2}$ as a function of the
    $r$ for the general Randall-Sundrum $6D$ metric with Weyl coordinates. The perturbed geodesics of the pure RS with Weyl coordinates is recovered when $N_{y1}=N_{y2}=0$.}
  \label{fig:kappa_RS6}
\end{figure}

\begin{figure}
  \centering
  \includegraphics[width=0.7\textwidth]{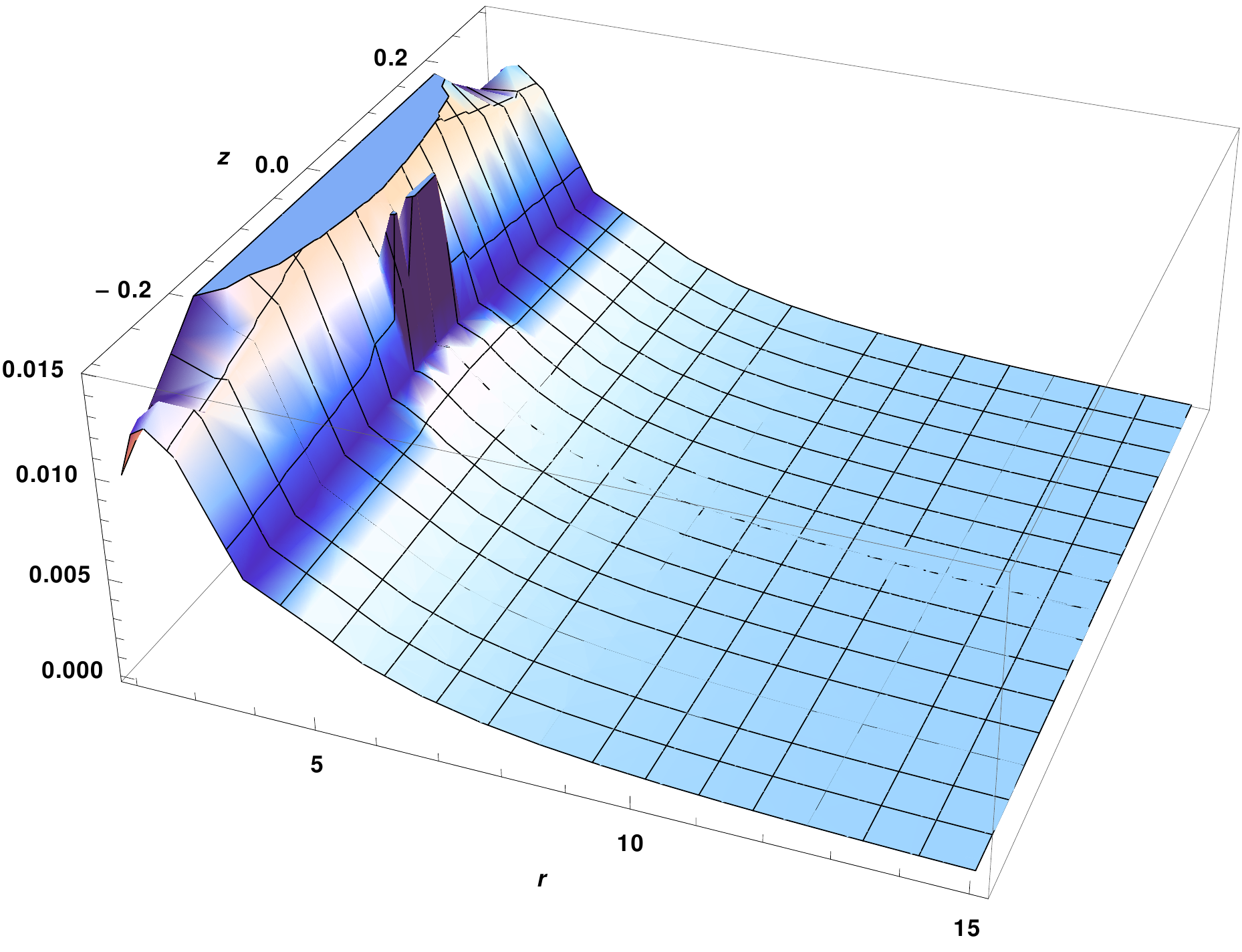}
  \caption{The plot shows $\tau_\perp^{2}$ as a function of
    $r$ and $z$ for the general Randall-Sundrum $6D$ metric with Weyl
    coordinates. Here it is assumed $N_{y1}=N_{y2}=0.3$, $a=0$ and and cut parameter is $c=1$ (satisfying energy conditions).}
  \label{fig:tau_RS6}
\end{figure}

Example 4 - general $6D$ Randall-Sundrum with axial symmetry
(Sec. \ref{ex4}): Fig. \ref{fig:kappa_RS6} and Fig. \ref{fig:tau_RS6}
show respectively the curves for $\kappa^2$ and $\tau_\perp^2$ for some
values of $N_{y1}$ and $N_{y2}$ in the case where the $6D$ axisymmetric
system is Randall-Sundrum. The system is stable if $\kappa^2>0$ and
$\tau_\perp >0$. When $N_{y1}=N_{y2} \rightarrow 0$, the curves are
stable both for the radial and orthogonal perturbations if $r\gtrsim
0.6$ (this is the pure RS system). The results show that any values of
$N_{y1}$ and $N_{y2}$ give stable results. In fact, there are a small
range of $r$'s (between 0 and $\sim 0.5$) that represents an unstable
region. This is a central region of a Miyamoto-Nagai gravitational
potential $\lambda = - \frac{m}{\sqrt{r^2 + (\bar{z}+a)^2}}$ (with $\bar{z}= |z| + c$), and such instabilities are indeed expected \cite{miyamoto}. The perpendicular perturbations of Fig. \ref{fig:tau_RS6} is plotted for this limit case when $N_{y1} = 0.20$ and $N_{y2} = 0.20$ and indicates that in the disk plane ($z\rightarrow 0$) the system is stable for all $r$. When $z$ acquires greater values, this range of stable radial regions is also greater. In this sense, one can conclude that two extradimensions carry instabilities but less instabilities than the $5D$ RS case. From $6D$ axisymmetric RS configuration, it is possible to conclude that {\bf two extradimensions does not carry instabilities for the system in the radial direction and also does not carry instabilities in the axial direction}. There are only local instabilities represented in Fig. \ref{fig:kappa_RS6} e.g. by the peaks around $r \approx 0.9$ ($N_{y1}=0.02$ and $N_{y2}=0.02$), around $r \approx 1.3$ ($N_{y1}=0.05$ and $N_{y2}=0.10$) and around $r \approx 2.25$ ($N_{y1}=0.20$ and $N_{y2}=0.20$). 

In all situations we have introduced a cut method to generate a disk solution and energy conditions are satisfied for cut parameters $c \geq 1$. In all stable examples cited above, the $\kappa$ frequency radial distribution follows what is expected for general relativity, i.e., the epicyclic frequency does not increase monotonically inward in the radial direction, but rather reaches a maximum at certain radius and then falls to zero at the radius of marginally stable circular geodesic. The results of the present work are important to contribute to all astrophysical solutions that retrieve axisymmetric configurations living in a $nD$ universe. For example, the impact of extra $D$ perturbations at AGN disks, galaxies in general, accretion around stellar black holes, etc. Several complementary discussions about the implication of this both in astrophysics as in cosmology can be seen e.g. in \cite{braneworld2,braneworld3,braneworld4,coimbrasss,coimbrasss1}.

An important final observation is that stable solutions were found only for the $6D$ cases and this actually coincides with some arguments in favor of an even $nD$ e.g. as has been emphasized by diferent authors that the Huygens principle does not hold for odd $nD$ (see \cite{huygens1,huygens2}).

\section*{Acknowledgments}

The authors are very grateful to the researchers of DEE-UFPR. CHC-A specially thanks Patricio Letelier ({\it in memoriam}) who idealized the first steps of the present work. The authors are also very grateful to CQG referees for enlightening views and for suggestions given in order to improve the quality of this paper.


\vspace{5cm}

\end{document}